# A Lightweight Adaptable DNS Channel for Covert Data Transmission

Mahboubeh Nazari[1], Sousan Tarahomi[1*], Sobhan Aliabady[1]

[1] Department of Computer and Information Technology, Imam Reza International University, Mashhad, Iran
*s1.tarahomi@gmail.com

**Abstract:** Due to the vital role of security in online communications and this fact that attackers are developing their tools, modernizing the security tools is an essential. The efficiency of crypto systems has been proven after years, however one may need to communicate stealthy without drawing attention especially in transferring secret data such as keys. Covert channels are suitable tools that used to conceal the existence of data besides end communication parties by employing principles of steganography. They can make secure communications with obscurity. Working stealthy and providing an acceptable throughput are issues in designing covert channels. The DNS protocol properties like its necessity for running applications and the availability can provide aforementioned issues decently. In this paper, we proposed a storage covert channel which uses DNS protocol as a media for transferring data. The key features include connection establishment, adaptability with network environment, implying a lightweight obfuscation method and HMAC to meet confidentiality and integrity. Experimental results show the proposed channel statistics are well adapted with normal traffics. The channel has an average capacity of 2.65 bytes of data per packet.

1. **Introduction**

The term of Secure communication often linked with encryption or crypto systems [1,2]. Since power and cost in computing resources are becoming more efficient, attacks such as brute force increasingly threaten data security. However algorithm developers increase encryption key size in responses [1], this can impose overhead on processing resources. In encrypted communication, the intruder knows that there exist a data that flows between two specific endpoints. In other words, cryptography does not hide both the existence of data and identity of participants in the communication. When attackers are aware of communications, they can steal and fake the user credentials by sniffing, hijacking, cryptanalysis and other possible attack [2]. Steganography provides a higher level of security by hiding the existence of the message in such a way that the intruder cannot understand the difference between original and stego carrier [3,4].

Covert channels can be used as methods which are not only covering the insufficiency of encryption by applying principles of steganography but also hide both the secret data and method of communication [1]. Covert channel is not a new idea and first introduced by Lampson in the 1970 [5]. Initially covert channel was a term to describe the inadvertent data leakage between processes on a monolithic system, then the term used for communication channels which leak information in violation of system or process restrictions [1].

Covert channels inject secret data in normal and authorized traffic in such a way that do not raise the attention of the outside listener besides concealing of data existence and identity of the communication parties. Covert channels can be divided into two subcategories: 1- storage covert channel 2- timing covert channel. Storage covert channels use flaws in the implementation of network protocols to hide their secret message [6]. On the other hand, timing channels are created by manipulating time samples in the communication system. Therefore, advanced timing channels emulate the overt channel behavior that leads to less detectability than storage channels, but they suffer from lack of capacity [1, 7].

However, covert channels often considered as a threat to network security. Indeed, their usage for creating secure communications have been discussed in recent years. Calhoun Jr et al. proposed a covert authentication technique which uses 802.11 rate switching as a covert channel for authentication messages [8]. Sawicki and Piotrowski discussed a covert channel to authenticate access points, this channel uses beacon frames and timestamp fields and takes advantage of the least significant bits of these fields [9]. A covert channel which represented by Xie and Zhao have been used for identity authentication by using the intervals between two consecutive packets to embed identity tag [2].

All of the aforementioned methods are timing covert channels. However, they can afford obscurity, but suffer from some drawbacks like the need of synchronization for the both sides and also a very limited capacity of data, just few bits in most cases. Therefore, for an efficient channel, we need to employ storage covert channels, which leverage network protocols as a carrier for transferring data. Protocols such as ICMP, IP, and TCP have been used for storage channels [10]. As an example, a storage covert channel proposed by Muchene et al. to report insider threats, which benefits of using the differences in the framing approaches used by Ethernet and IP packets to append hidden information within IP packets and transmits it to an organization administrator [11]. Layer 7 (Application) protocols such as HTTP and DNS are also commonly used. Being prevalence on the Internet is not the main factor to select these protocols, standard practices show they are allowed through network protection devices by default and it makes them a suitable protocol for designing covert channels [10].

Domain Name System (DNS) is a significant protocol and service on the internet which is commonly used for mapping domain names into IP addresses. The ports are used for communication are UDP server port 53 and TCP server port 53. UDP is commonly used, but TCP will be used for zone transfers or with payloads over 512 bytes [12].

Generally, we have two common issues in designing covert channels, channel capacity and detectability. Since DNS is a prevalence, bidirectional, recursive protocol and it is necessary for running applications, it can be obscure as a covert channel. In addition, DNS covert channels can achieve a bandwidth of 110 kilobytes per second with 150 ms of latency [13], which is an acceptable capacity for a storage channel.

The key features of DNS make it a suitable protocol for creating covert channels. These channels can be used for a full remote control channel, capabilities including Operating System (OS) commands, file transfers, full IP tunnel or penetration testing tool [13, 14].

DNS covert channels can also come with DNS tunneling term. In DNS tunneling, another protocol is tunneled with DNS. Pearson first discussed DNS covert channels in 1998. All of them have similar basis, but they come with different encoding and implementation details. Main techniques include a controlled domain or subdomain, a server side entity, a client side entity and encoded data in DNS payloads. The server is the authoritative name server for the controlled domain and responds to client queries for this domain. The client is the other end of the channel, which needs to query the server to send and receive data. However, DNS channels have existed since 1998, tunneling data over DNS widely presented in 2004 with proposing OzymanDNS by Kaminsky. Different record types and encoding methods have been used by DNS covert channel, but TXT and NULL are the ones with most used [13, 14, 15].

NSTX tunnels IP over DNS by splitting the IP packets into several DNS packets besides using of TXT records [16]. OzymanDNS can transfer TCP or SSH over DNS and supports of EDNS0, this tool transfer data in TXT records [17]. The idea of DNScat was started by NSTX which tunnels IP over DNS, but they are dissimilar in design and implementation. It uses A and CNAME records and provide a bidirectional communication through DNS servers [18]. DNS2TCP transfers TCP or SSH over DNS in TXT records [19]. TUNS connects a client in a firewalled network with a server on the internet. It splits IP packets using IP fragmentation in operating system and transfer them in CNAME records [15]. Heyoka encodes data in the hostname of queries for transferring data from client and TXT, NULL records have been used in this method [20]. Iodine is a tool proposed by Andersson in 2010, which transfers IP over DNS in a set of queries like NULL, TXT, SRV, MX, CNAME and A records. It needs to use a virtual interface (TUN/TAP) to make the channel [21]. However, Iodine transfer data in plain text, DNScat2 developed more advanced method for data security such as encryption, authentication and signature to provide confidentiality and integrity for data [22]. Most of the discussed methods, tunnel another protocol in DNS packets. We can transfer different data through DNS packets like botnets such as feederbot [23], which uses DNS as a command and control protocol.

In this paper we proposed a DNS covert channel to transfer data and meet requirements such as acceptable capacity and undetectablitiy in addition to compatibility with its network environment. Data in this channel is secure without imposing overhead by implying security methods. The rest of the paper organized as follows: in section 2, the proposed scheme for covert channel are discussed in more details. After that, experimental results and analysis of the proposed channel provided in section 3. Finally, in section 4, we come into conclusion.

## 2. Proposed scheme

The channel properties and the operation have been explained in this section. Sections 2.1 to 2.7 are about channel properties and data transmission is discussed in section 2.8.

### 2.1. Data Encapsulation

DNS works on UDP as a connectionless protocol with no guarantee for packet delivery. Packets need to be traced in case of packet lost. Therefore, we benefits of TCP principles in data encapsulation. The sequence numbers are used to track packets and prevent of retransmissions. In contrast to TCP, only one sequence number is used in both sides to save more space for data and make the channel simple. The structure is shown in Fig. 1.

| Flags | ID | Sequence number | Message | HMAC |
|---|---|---|---|---|

***Fig. 1.*** *The packet structure in the proposed channel*

Flags: they are used for control purposes. Three bits are dedicated to flags, one bit per each flag. Flags are defined as below:

Data_type: Both of communication parties need to know the type of data in order to decide how to behave with it. The data type is distinguished into text and binary. Before data transmission, its type is checked and the Data_type flag will be set. The 0 value is for the binary data and 1 for the text data.

False integrity: it checks the integrity of data by checking the HMAC part. When the HMAC code of a packet is incorrect, the receiver of packet sets this flag to 1 and sends another packet. The sequence number in this packet is supposed to be the same number as corrupted packet. When the sender receives such packet, it will notice there was a problem in the packet and resend it.

Fin: each side sets this flag to 1 while they have no data to send.

ID: a one byte random number as the client identification. Each client will get a unique ID.

Sequence Number: a two byte random number to represent the packets sequence. The client selects the initial sequence number (ISN) and the number is incremented one by one for each packet.

Message: the data needs to be sent. It can be a file, string, etc.

HMAC: the HMAC code is calculated to check the integrity of data and then attached to the end of the packet.

### 2.2. Candidate Record Type Selection

The covert channel is supposed to work in a network, since this environment is constantly changing, the network condition needs to be analysed to adapt the channel activity with its working network. Thus, before the channel starts working, a real time process checks the network and DNS record types with more frequency in the current traffic will be selected. These record types are considered as candidate records and will be stored. Whenever the channel has data to transfer, one candidate record type will be selected randomly. This approach makes the channel obscure and adaptable with its network.

### 2.3. Encoding Type

In order to prevent raise of attention to the channel, queries need to look normal like standard DNS packets. Therefore, Base32 is used to encode queries from the client. Base32 encoding is designed to represent arbitrary sequences of octets in a form to be case insensitive, but not human readable [24]. Base32 or 5-bit encoding is commonly used for requests from the client [13] and also references such as DNScurve [25] and RFC5155 [26] use Base32 encoding in queries for security purposes. Base32 utilization as an encoding method in one hand leads to lower capacity (5 bits per character), but on the other hand it makes the channel more adapted with normal traffics. DNS queries and responses pass through the ISP's DNS infrastructure, they must not be too different with normal DNS packets. The responses in record types such as CNAME which have a similar response structure with requests also encode in Base32. In order to encode more data, Base64 or 6-bit encoding is used for TXT record responses.

### 2.4. Variable Packet Length

Some detection methods, like Guy, consider DNS queries and response length, which look at all hostname requests longer than 52 characters. DNS covert channels usually try to embed as much as possible data in requests and response, therefore, they will have long labels up to 63 characters and long overall names up to 255 characters [13].

Based on RFC1034 [27], the hostname in DNS query is limited to 255 characters and it is structured in labels. Labels must be 63 characters or less. In this implementation, test.com is considered as the channel domain. For this domain, the hostname with maximum amount of data would be look like this:

63chars.63chars.63chars.54chars.test.com

In encoding with Base32, The maximum amount of data would be 1215 bits (151 bytes) per packet, definitely in Base64 this amount will be more. In order to provide more reliability and prevent of recognition by length based detection methods, we don't embed data to the maximum amount. After encoding data in Base32, the volume of data decreases because of 5 bits per character, in order to not make large packets length and payload, a range with a maximum of 10 and a minimum of 2 bytes is considered. Whenever a query needs to be sent, a random number in range (2, 10) will be selected as the total amount of message, also the labels length will be changed randomly. This range can be modified based on the statistics in network evaluation phase as mentioned in section 2.2. It means, if most of the normal packets in the network are large, bigger numbers will be selected for the range, otherwise the range will have small numbers like this current implementation.

This method works for upstream data (client to server) which the data is always encapsulated in the hostnames. It can also work for downstream data (server to client) with record types such as CNAME, which have a similar response structure with requests. For other record types in downstream like TXT, we can only change the total amount of encapsulated data. The header length of encapsulated data is invariable for all packets, but with altering the amount of message, the total length of packets will change and this approach prevents of having uniformed packets, which can be suspicious.

### 2.5. A Lightweight Obfuscation Method to Obscure the Data Stream

The data in channel needs to get defaced to prevent of leaking information even after the channel gets discovered. However, Iodine [21] transfers data in plain text and Feederbot just uses a simple encryption method like RC4 [23], Dnscat2 benefits from some stronger encryption methods like salsa20.

In the proposed channel, we just need to deface the data without imposing more overhead on the channel. Therefore the data get obfuscated not encrypted, which needs key exchange. It is assumed there is a pre-shared secret between communication parties. A pseudorandom number generator (PRNG) is used. PRNG seed is generated based on (1):

$$seed = Preshared\_secret \oplus ISN,$$
$$R_i = random(seed) \quad (1)$$

Where ISN is the initial sequence number has exchanged in the beginning of the connection establishment. ISN number is supposed to be unique for each client. Whenever a packet needs to be sent, Ri will be generated and i is the number of the generated packet. The data is get obfuscated based on (2).

$$Obfuscate\_data = data \oplus R_i \quad (2)$$

After obfuscation, the data will be encoded in Base32 or Base64 and encapsulate in the packet.

### 2.6. Integrity Check

Feederbot uses a simple CRC32 code to check the data integrity and Dnscat2 signs the data. In the proposed channel, data integrity is provided by HMAC code.

Hence we want to use HMAC only for integrity check and we rely on obfuscation for security purposes. MD5 is used as the hash function for HMAC. MD5 has the shorter output with 128 bits in comparing with SHA family, and shorter output gives us the opportunity to transfer more data per packet. The HMAC key generation is stated in (3).

$$Key = MD5(Preshared\_secret || ISN) \quad (3)$$

And the HMAC function works like (4):

$$HMAC = MD5(data||ISN||Preshared\_secret) \quad (4)$$

The HMAC code is added after the encoded data as discussed in section 2.1. The data in HMAC part is the message before obfuscation and encoding, it doesn't include header which has flags, ID and sequence number. The header needs to transfer in plain text for control purposes. When a packet is received, first the data is decoded and deobfuscated, then the message is obtained. The other party has to generate the HMAC code and compares it with the received code, if they are equal it means the data has transferred properly. Otherwise the false integrity flag will be set to 1 and a packet will be sent with the same sequence number and an arbitrary string in the message section. When the sender of packet receives a packet with the same sequence number as its previous packet has just sent, it will notice there was a problem in that packet and resend it.

### 2.7. Packet Loss and Reordering

For being undetectable, DNS queries in the proposed channel pass through the pre-configured DNS resolver on the host and unlike Feederbot, the channel does not query the server directly. Bypassing the pre-configured DNS resolver can be suspicious. Therefore the pre-configured DNS server on the host is responsible for receiving the replies and our DNS queries will behaved like other queries in the host. In the server side, since there is one ID per client, until the server doesn't receive a DNS query from the client with Fin flag 1, receiving packets will continue, and the server keeps the connection open. The client can encounter with a problem and the server doesn't receive any query from the client for a long time, the server needs to terminate the connection one sided.

For this approach, a time interval is set based on normal TTL values for DNS records. If this interval is expired and the server had not received any packet from the client, the server terminates the connection and the client ID will be removed from the server side. If the client connects the server after this time expiration, the server considers this connection as a new one.

As sequence numbers are utilized in the channel, both sides have to discard packets which have an equal or smaller sequence number with the last packet they received, for server this is done per each ID. There is an exception has explained in section 2.6 that packets with the same sequence number won't be discarded.

### 2.8. The Communication Phases

This section is about the phases in which client and server as two sides of covert channel start a communication, transfer the data and terminate the connection after they finish sending data.

*2.8.1 Connection Establishment:* The proposed channel works on DNS, the client needs to start the communication and the server responses to the client request. First the client generates one byte random number as its ID and a two bytes random number as initial sequence number (ISN). Encapsulated data will be structured as explained in section 2.1.

*2.8.2 Data Transmission from Client Side*: Data transfer from the client side will be done in the following steps:

1. Before sending data, its data type will be checked and the Data_type flag will be set to an appropriate value.
2. The amount of message which needs to transfer is selected randomly in the given range as described in section 2.4
3. The message gets obfuscated and encoded with the header part based on the method discussed in section 2.5 and the HMAC code is added as explained in section 2.6.
4. The record type is selected randomly among of candidate record types as described in section 2.2.
5. Data from the client is always transferred as a request, therefore we need to embed the data in the hostname for example: data_from_client.test.com.

If there is no data left for sending but the server still have data, the client sets its Fin flag to 1 and it indicates the client finished sending data. Then a random string is generated and encapsulated in the message part. The random string with flags, ID and sequence number follow steps like data packets. Then, it will be encapsulated in the hostname of the query. The proposed scheme is shown in Fig. 2.

*2.8.3 Data Transmission from Server Side:* When a request in defined domain (for example: test.com) receives, first the server decodes and deobfuscates the data as (5), (6):

$$Obfuscated\_data = Base32.decode(Recieved\_data) \quad (5)$$

$$R_i = random(seed), data = Obfuscated\_data \oplus R_i \quad (6)$$

If the server has data to send, it follows the steps 1 to 3 as described above in section 2.8.2. Data from the server is embedded in rdata part of the response. In case of no data to send, since response to a DNS request is compulsory, whether there is data in the client side or not, the value of Fin flag is set to 1 to demonstrate the server has no data to send. Then a random string is generated and placed in the message part. The random string with flags and ID and sequence number follow the steps like data packets, then encapsulated in rdata part of the response and will be sent. The proposed scheme is shown in Fig. 3. Steps in Fig.3 are the same ones in Fig. 2.

*2.8.4 Connection Termination:* When each part of the communication finish sending data, they set the fin flag value to 1, besides a random string will be located in the message part and encapsulate with flags, ID and sequence number in a packet. Data receiving will continue until completion of data transfer by the other side, which will be determined by setting the fin flag to 1. When one side is about to terminate the connection, it needs to check whether the other side has declared finishing data or not. While the fin flag has set to 1 in both sides, each of parties can terminate the connection.

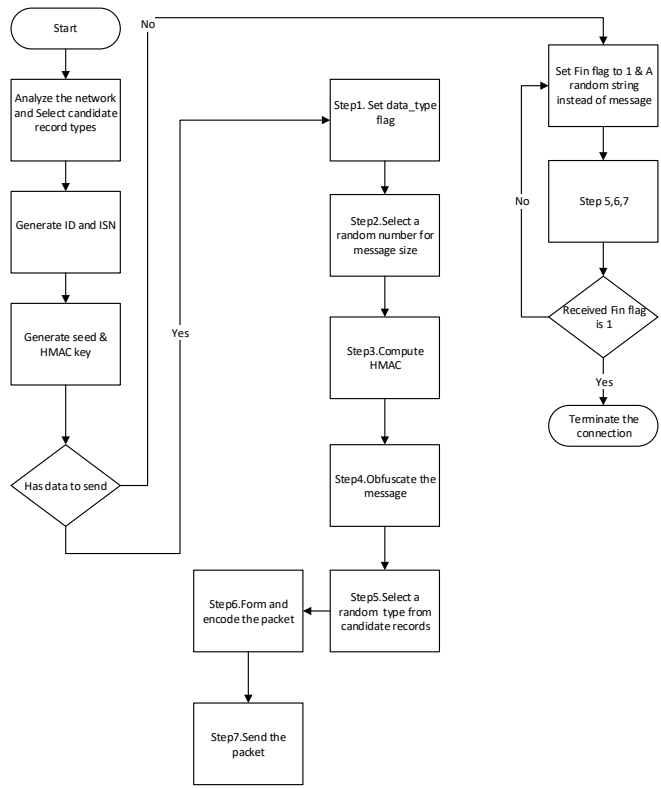

*Fig. 2. Data transmission from the client side*

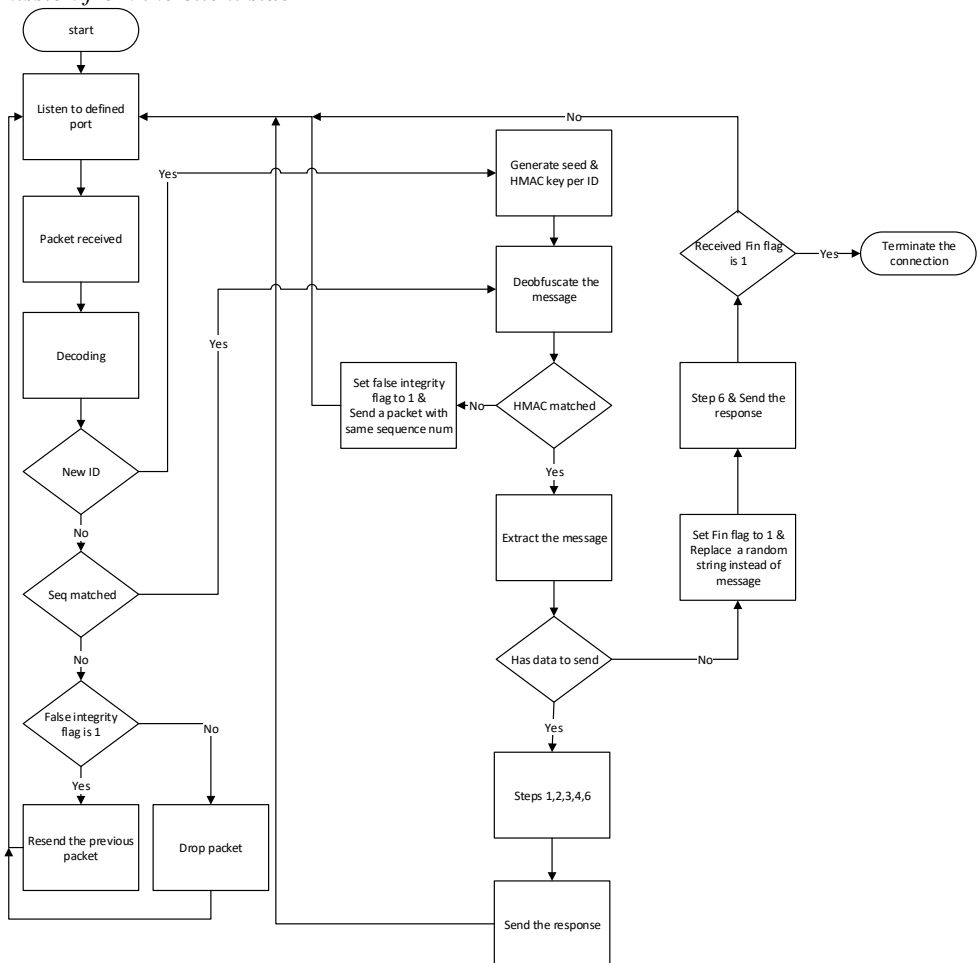

*Fig. 3. Data transmission from the server side*

## 3. Evaluation and results

The proposed channel is implemented in python programing language. The candidate record types are A, CNAME and TXT in this implementation. In order to check its adaptability with normal traffics, the proposed channel statistics were compared with two datasets of normal network traffic.

Dataset A: a normal capture of traffic in a home network. Capture duration is 1.9 hours and the number of packets is 5966. The performed actions are included: - Music streaming from 20songstogo.com - Gmail - Twiter - Jitsi chat connected to gtalk, SIP and CVUT jabber. - Some ssh - cacti web - normal webs with chrome. The capture has done in 2015-03-24.

Dataset B: a normal capture in a Linux Debian notebook computer in an xDSL network. Capture duration is 1281.885508 seconds (21 minutes). Applications and actions in the normal computer are: Deluge P2P on linux notebook. Downloading some large files. The p2p was running for 1 hour when the capture was started. At the beginning there is also an mtr sending icmp packages to www.google.com .Some web pages were accessed: twitter, facebook, etc. All using the Chrome browser. The capture has done in 2013-12-17.

In addition to these datasets, we compared the proposed channel with Dnscat2. Since other previous DNS channels as named in the introduction section are not new or have different purposes, we compare our channel with Dnscat2 as the latest one. Dnscat2 includes many options, since the proposed channel is designed only for data transfer, data transferring in Dnscat2 has been compared.

Files with different random sizes were used as the message, then they were transferred by both of the proposed channel and Dnscat2. Tests had been done in a local network with two virtual machines as client and server. The operating system is Kali-Linux-2016.1 in the server and Ubuntu-14.04.2 in the client.

### 3.1. Length and Size

Table 1 provides the statistics for Datasets A and B that gained by Wireshark. The proposed channel was evaluated in payload size, qname length and packet length with Datasets A, B as normal DNS traffics to check the similarity with normal traffics. Also the channel is compared with Dnscat2.

Tables 2 and 3 show the statistics for 12 times running proposed channel and Dnscat2 with different file sizes as exchanged data. Results show the minimum value for payload size, qname length and packet length are constant in Dnscat2, while in proposed channel, both minimum and maximum values are variable as defined in section 2.4. The Average value of payload size in the proposed channel and Dnscat2 are 68.28 and 94.78, respectively. It is noticeable that values of proposed channel are more compatible with average values in Dataset A with 71.65 and Dataset B with 70.62 rather than Dnscat2. The Average value of packet length in the proposed channel is 110.28 and for Dnscat2 is 136.78. This results prove that our channel is more adaptable with Datasets A and B than Dnscat2 in the similar circumstance. Average qname length is 37.71 in the proposed channel and 50.93 in Dnscat2. This value is 19.94 and 27.48 for Dataset A and Dataset B respectively. However, qname lengths in the proposed channel is shorter than Dnscat2 qname lengths, still there is a difference with normal traffics. Base32 is used in encoding records, which cause longer lengths, but this doesn't make the queries abnormal. Also the maximum qname length in our proposed method is 46 and 47, and the maximum values for Datasets as normal DNS traffic are 44 and 36. Therefore, the proposed channel qname length is adaptable with maximum values of normal traffics and this difference won't be noticeable in huge amount of DNS traffics which are transferring daily.

### 3.2. Standard Looking Queries

As Table 4 shows most of queries in normal traffics fall into 4th level or 3rd level. In the proposed channel all of the queries fall into 4th level, but in Dnscat2 the majority are in 3rd level. This result indicates that queries in the proposed channel do not differ with normal queries and follow the normal standards.

**Table 4** Label statistics for queries in datasets A and B

| Dataset | 4th Level or More | 3rd Level | 2nd Level | 1st Level |
|---|---|---|---|---|
| A | 253 | 2575 | 197 | 4 |
| B | 1007 | 60 | 7 | 1 |

**Table 1** Statistics for datasets A and B

| Dataset | Packet count | Payload size | | | Qname len | | | Packet len | | |
|---|---|---|---|---|---|---|---|---|---|---|
| | | min | max | avg | min | max | avg | min | max | avg |
| A | 5966 | 22 | 377 | 71.65 | 6 | 44 | 19.94 | 64 | 419 | 113.56 |
| B | 2147 | 27 | 262 | 70.62 | 11 | 36 | 27.48 | 69 | 304 | 112.62 |

**Table 2** The statistics for proposed channel and Dnscat2 in transferring 12 different data sizes

| Total transferred bytes | Payload size | | | | | | Qname length | | | | | |
|---|---|---|---|---|---|---|---|---|---|---|---|---|
| | Proposed channel | | | Dnscat2 | | | Proposed channel | | | Dnscat2 | | |
| | min | max | avg | min | max | avg | min | max | avg | min | max | avg |
| 12 | 53 | 95 | 67.17 | 61 | 346 | 93.28 | 37 | 38 | 37.33 | 45 | 159 | 48.75 |
| 19 | 53 | 107 | 72.63 | 61 | 346 | 89.78 | 37 | 46 | 39.75 | 45 | 159 | 46.14 |
| 23 | 53 | 105 | 70.20 | 61 | 183 | 90.27 | 37 | 40 | 37.80 | 45 | 108 | 47.03 |
| 30 | 50 | 102 | 67.83 | 61 | 197 | 95.36 | 34 | 37 | 36 | 45 | 122 | 50.50 |
| 41 | 48 | 107 | 69 | 61 | 218 | 92.66 | 32 | 39 | 37.38 | 45 | 145 | 49 |
| 52 | 46 | 107 | 69.33 | 61 | 242 | 93.02 | 30 | 46 | 37.11 | 45 | 167 | 48.94 |
| 69 | 50 | 103 | 68.50 | 61 | 265 | 93.06 | 34 | 47 | 38.07 | 45 | 202 | 49.62 |
| 76 | 44 | 97 | 66.36 | 61 | 279 | 93.77 | 28 | 40 | 36.36 | 45 | 216 | 50.18 |
| 87 | 46 | 110 | 67.56 | 61 | 313 | 94.67 | 30 | 46 | 37.06 | 45 | 238 | 51.66 |
| 93 | 46 | 96 | 64.84 | 61 | 313 | 100.42 | 30 | 46 | 36.94 | 45 | 240 | 56.39 |
| 100 | 50 | 103 | 69.80 | 61 | 315 | 103.90 | 34 | 46 | 40.60 | 45 | 240 | 59.60 |
| 112 | 46 | 103 | 66.11 | 61 | 313 | 97.21 | 30 | 46 | 38.11 | 45 | 240 | 53.41 |

**Table 3** Packet length for proposed channel and Dnscat2 in transferring 12 different data sizes

| Total transferred bytes | Proposed channel | | | DNScat2 | | |
|---|---|---|---|---|---|---|
| | min | max | avg | min | max | avg |
| 12 | 95 | 137 | 109.17 | 103 | 388 | 135.28 |
| 19 | 95 | 149 | 114.63 | 103 | 388 | 131.78 |
| 23 | 95 | 147 | 112.20 | 103 | 225 | 132.27 |
| 30 | 92 | 144 | 109.83 | 103 | 239 | 137.36 |
| 41 | 90 | 149 | 111.00 | 103 | 260 | 134.66 |
| 52 | 88 | 149 | 111.33 | 103 | 284 | 135.02 |
| 69 | 92 | 145 | 110.50 | 103 | 307 | 135.06 |
| 76 | 86 | 139 | 108.36 | 103 | 321 | 135.77 |
| 87 | 88 | 152 | 109.56 | 103 | 355 | 136.67 |
| 93 | 88 | 138 | 106.84 | 103 | 355 | 142.42 |
| 100 | 92 | 145 | 111.80 | 103 | 357 | 145.90 |
| 112 | 88 | 145 | 108.11 | 103 | 355 | 139.21 |

### *3.3. Security Consideration*

HMAC code is applied to check the integrity of data, so data corruption would be detectable in exchange. The channel can meet data integrity besides error detection by HMAC code utilization. The data obfuscation is implied not only to deface the data, but also to make it unreadable by an intruder, which satisfies confidentiality properties. In order to not impose a more overhead on the channel, any kind of encryption and authentication schemes are not used in the proposed scheme. Instead, a pre-shared value is used as the secret parameter to build a shared key between the communication parties.

### *3.4. Channel Capacity*

Table 5 displays the amount of data as the number of packets which are needed to complete data transfer. In order to meet adaptability and prevent of making abnormal traffic, the encapsulated data in a single packet is not high, however this option can be changeable due to channel environment as explained in section 2.4. For this implementation and with these current parameters, the average capacity of channel would be 2.65 bytes of data per packet. The only overhead in data encapsulation is because of double issues. First, the header needs to repeat in every packet and the second is the encoding type as discussed in section 2.3.

Previous channels such as Dnscat2, keep their session alive with constantly pulling the server, in result they inject many DNS packets in network traffic. In our examination, 72 packets were exchanged in 30 seconds running Dnscat2. This issue can make anomaly and overhead in the network even for transferring a low amount of data. In contrast, the client in the proposed channel doesn't have to pull the server for keeping the connection alive, because the channel is implied for data transmission and the connection will be terminated after it.

The proposed channel capacity is acceptable since there is no extra packets as pulling ones and the packets just transfer data without imposing overhead on the network.

Furthermore, some confidential data such as secret keys are not big, they can be transferred with the proposed channel that provides them obscurity while exchange.

**Table 5** Channel capacity

| Total transferred bytes | Packet count | Bytes per packet |
|---|---|---|
| 12 | 6 | 2 |
| 19 | 8 | 2.37 |
| 23 | 10 | 2.3 |
| 30 | 12 | 2.5 |
| 41 | 16 | 2.56 |
| 52 | 18 | 2.88 |
| 69 | 28 | 2.46 |
| 76 | 28 | 2.71 |
| 87 | 32 | 2.71 |
| 93 | 32 | 2.90 |
| 100 | 30 | 3.33 |
| 112 | 36 | 3.1 |

### 3.5. Detectability

For evaluating channel detectability, the proposed channel was tested with Opnsense 17.1 as an open source SPI firewall and Suricata 3.2.1 as an open source IDS. Suricata rules have been updated to the date of 2017-09-28. Two test scenarios were designed, one with Opnsense and another for Suricata.

As the client and server are running in virtual machines, Opnsense firewall was installed on another virtual machine and set as a gateway between the client and the server. All traffics from the client to the server and reverse pass through the firewall. The firewall was configured to monitor both incoming and outgoing traffics. After running Opnsense, the channel started running and transferring files. The firewall didn't record anything suspicious about the channel.

In the next scenario, the channel detectability had been tested by an IDS and in our case Suricata was used. Since in DNS covert channels scenarios is assumed the server is under control and the client is the machine which is not in control, so the IDS was installed on the client side to check channel detectability by IDS updated rules. Similar to the previous scenario, after running Suricata, the channel started working and transferring files. To make an accurate estimate, the tests were repeated for varied data sizes as files. The IDS didn't record any suspicious activity.

### 3.6. RFC Compliant

In this section, the proposed channel is evaluated with the following items to show the channel compatibility with standards in RFCs.

*3.6.1 Queries and Responses Size:* The proposed covert channel doesn't proceed the limited size as mentioned in RFCs. While most of covert channels try to put as much as data in the packets, the proposed channel tries to make a good balance between undetectability and capacity.

*3.6.2 Common Record Types*: There is an evaluation phase before the channel starts working and this leads to selecting common records in current traffic. This approach prevents of selecting uncommon records that makes the channel detectable.

*3.6.3 Encoding Type:* The encoding methods were selected for encoding queries and responses are RFC compliant and commonly used.

### 4. Conclusion

We proposed a covert channel based on DNS protocol for covert data transmission. The channel can get adaptable with its network environment and works stealthy. The channel works with common DNS record types, encoding, and it is RFC compliant. A lightweight obfuscation method has implied for data encapsulation without imposing overhead on the channel. This method is more lightweight in comparison with Dnscat2, which needs key exchange and encryption. Data integrity in the channel has been provided by HMAC. Experimental results show the channel has a good compatibility with normal traffics. The evaluation with two security software, Opnsesne as an SPI firewall and Suricata as an open source IDS with latest updated rules, didn't record anything suspicious. The channel capacity is an average 2.65 bytes of data per packet, which is acceptable because of the encoding type and undetectablility. We can also improve the capacity by adding another phase to evaluate the packet sizes of the network and set the channel parameters based on them. This option can lead to a higher capacity without drawing attention. The channel presented in this paper has a good capability for transferring confidential data such as key exchange, user credentials and other secret information.